\documentclass[runningheads]{llncs}
\usepackage{graphicx}
\usepackage{multirow}
\usepackage{graphicx}
\usepackage{amsmath}
\usepackage{multirow}
\usepackage{booktabs}
\usepackage{array}
\usepackage{subfigure}
\usepackage{amssymb}
\usepackage{algorithm}
\usepackage{verbatim}
\usepackage{bm}     
\usepackage{enumitem}       
\usepackage{lipsum}
\usepackage{titlesec}

\titlespacing*{\subsection}
{0pt}{2pt}{1pt}
\titlespacing*{\subsubsection}
{0pt}{0pt}{2pt}

\setenumerate[1]{itemsep=0pt,partopsep=0pt,parsep=\parskip,topsep=0pt,itemindent=0pt}
\setitemize[1]{itemsep=0pt,partopsep=0pt,parsep=\parskip,topsep=0pt,itemindent=0pt}

\begin{document}

\title{Multi-view Multi-behavior Contrastive Learning in Recommendation}

\author{Yiqing Wu$^{\dagger}$\inst{1,2,3}\and Ruobing Xie$^{\dagger}$\inst{3} \and Yongchun Zhu\inst{1,2}\and Xiang Ao\inst{1,2}\and Xin Chen\inst{3}\and Xu Zhang\inst{3}\and Fuzhen Zhuang\inst{4,5} \and Leyu Lin\inst{3} \and Qing He\thanks{indicates corresponding author. $\dagger$ indicates equal contributions.}\inst{1,2}}
\institute{Key Lab of Intelligent Information Processing of Chinese Academy of Sciences (CAS), Institute of Computing Technology, CAS, Beijing 100190, China
\and University of Chinese Academy of Sciences, Beijing 100049, China
\and WeChat Search Application Department, Tencent, China
\and Institute of Artificial Intelligence, Beihang University, Beijing 100191, China
\and Xiamen Institute of Data Intelligence, Xiamen, China\\
\email{\{wuyiqing20s,zhuyongchun18s,aoxiang,heqing\}}@ict.ac.cn,\\
\email{\{ruobing xie,andrewxchen,xuonezhang,goshawklin\}@tencent.com}, \email{zhuangfuzhen@buaa.edu.cn}}

\maketitle              
%

\begin{abstract}
Multi-behavior recommendation (MBR) aims to jointly consider multiple behaviors to improve the target behavior's performance. We argue that MBR models should: (1) model the coarse-grained commonalities between different behaviors of a user, (2) consider both individual sequence view and global graph view in multi-behavior modeling, and (3) capture the fine-grained differences between multiple behaviors of a user. In this work, we propose a novel Multi-behavior Multi-view Contrastive Learning Recommendation (MMCLR) framework, including three new CL tasks to solve the above challenges,  respectively. The multi-behavior CL aims to make different user single-behavior representations of the same user in each view to be similar. The multi-view CL attempts to bridge the gap between a user's sequence-view and graph-view representations. The behavior distinction CL focuses on modeling fine-grained differences of different behaviors. In experiments, we conduct extensive evaluations and ablation tests to verify the effectiveness of MMCLR and various CL tasks on two real-world datasets, achieving SOTA performance  over existing baselines. Our code will be available on \url{https://github.com/wyqing20/MMCLR}

\keywords{multi-behavior recommendation \and contrastive learning}
\end{abstract}

\section{Introduction}

Personalized recommendation aims to provide appropriate items for users according to their preferences. The core problem of personalized recommendation is how to accurately capture user preferences from user behaviors. In real-world scenarios, users usually have different types of behaviors to interact with recommender systems. For example, users can \emph{click}, \emph{add to cart}, \emph{purchase}, and \emph{write reviews} for items in E-commerce systems (e.g., Amazon, Taobao).
Some conventional recommendation models~\cite{sun2019BERT4Rec} often rely on a single behavior for recommendation. However, it may suffer from severe data sparsity~\cite{singh2008relational,pan2010transfer,zhu2021personalized} and cold-start problems~\cite{pan2019warm,xie2020internal,zhu2021transfer,zhu2021learning} in practical systems, especially for some high-cost and low-frequency behaviors . In this case, other behaviors (e.g., \emph{click} and \emph{add to cart}) can provide additional information for user understanding, which reflect user diverse and multi-grained preferences from different aspects. 


\textbf{Multi-behavior recommendation (MBR)}, which jointly considers different types of behaviors to learn user preferences better, has been widely explored and verified in practice \cite{chen2020efficient,chen2021graph,xi2021modeling}.
ATRank \cite{zhou2018atrank} uses self-attention to model feature interactions between different behaviors of a user in sequence-based recommendation with focusing on the \textbf{individual sequence view} of a single user's historical behaviors. In contrast, MBGCN \cite{jin2020multi} considers different behaviors in graph-based recommendation, focusing on the \textbf{global graph view} of all users' interactions.
However, there are still three challenges in MBR:


(1) \textbf{\emph{How to model the coarse-grained commonality between different behaviors of a user?}}
All types of behaviors of a user reflect this user's preferences from certain aspects, and thus these behaviors naturally share some commonalities. Considering the commonalities between different behaviors could help to learn better user representations to fight against the data sparsity issues. However, it is challenging to extract informative commonalities between different behaviors for recommendation, which is often ignored in existing MBR models.

(2) \textbf{\emph{How to jointly consider both individual and global views of multi-behavior modeling?}}
Conventional MBR models are often implemented on either sequence-based or graph-based models separately based on different views. The sequence-based MBR focuses more on the individual view of a user's multiple sequential behaviors to learn user representations \cite{zhou2018atrank}. In contrast, the graph-based MBR often concentrates on the global view of all users' behaviors, with multiple behaviors regarded as edges \cite{jin2020multi}. Different views (individual/global) and modeling methods (sequence/graph-based) build up different sides of users, which are complementary to each other and are helpful in MBR.

(3) \textbf{\emph{How to learn the fine-grained differences between multiple behaviors of a user?}}
Besides the coarse-grained commonalities, users' multiple behaviors also have fine-grained differences. There are preference priorities even among the target and other behaviors (e.g., \emph{purchase} $>$ \emph{click}).
In real-world E-commerce datasets, the average number of \emph{click} is often more than $7$ times that of the average number of \emph{purchase} \cite{jin2020multi}. The large numbers of clicked but not purchased items, viewed as hard negative samples, may reflect essential latent disadvantages that prevent users to purchase items. Existing works seldom consider the differences between multiple behaviors, and we attempt to encode this fine-grained information into users' multi-behavior representations.

Recently, contrastive learning (CL) has shown its magic in recommendation, which greatly alleviates the data sparsity and popularity bias issues \cite{zhou2020s3}. We find that CL is naturally suitable for modeling coarse-grained commonalities and fine-grained differences between multi-behavior and multi-view user representations. To solve above challenges, we propose a novel \textbf{Multi-behavior Multi-view Contrastive Learning Recommendation} (\textbf{MMCLR}) framework.
Specifically, MMCLR contains a sequence module and a graph module to jointly capture multiple behaviors' relationships,  learning multiple user representations from different views and behaviors. We design three contrastive learning tasks for existing challenges, including the multi-behavior CL, the multi-view CL, and the behavior distinction CL.
(1) The \emph{multi-behavior CL} is conducted between multiple behaviors in both sequence and graph views. It assumes that user representations learned from different behaviors of the same user should be closer to each other compared to other users' representations, which focuses on extracting the commonalities between different types of behaviors.
(2) The \emph{multi-view CL} is a harder CL conducted between user representations in two views. It highlights the commonalities between the local sequence-based and the global graph-based user representations after behavior-level aggregations, and thus improves both views' modeling qualities.
(3) The \emph{behavior distinction CL}, unlike the multi-behavior CL, concentrates on the fine-grained differences rather than the coarse-grained commonalities between different types of behaviors. It is specially designed to capture users' fine-grained preferences on the target behavior's prediction task (e.g., purchase).
The combination of CL tasks can multiply the additional information brought by multiple behaviors in the target recommendation task. Through the MMCLR framework assisted with three types of auxiliary CL losses, MBR models can better understand the informative commonalities and differences between different user behaviors and modeling views, and thus improve the overall performances.

In experiments, we evaluate MMCLR on real-world MBR datasets. The significant improvements over competitive baselines and ablation versions demonstrate the effectiveness of MMCLR and its different CL tasks and components. The contributions of this work are summarized as follows:
\begin{itemize}
  \item We systematically consider multiple contrastive learning tasks in MBR. To the best of our knowledge, this is the first attempt to bring in contrastive learning in multi-behavior recommendation.
  \item We propose a multi-behavior CL task and a multi-view CL task, which model the coarse-grained commonalities between different behaviors and (individual sequence/global graph) views for better representation learning.
  \item We also design a behavior distinction CL task, which creatively highlights the fine-grained differences and behavior priorities between multiple behaviors via a contrastive learning framework.
  \item MMCLR outperforms SOTA baselines on all datasets and metrics. All proposed CL tasks and the capability on cold-start scenarios are also verified.
\end{itemize}

\section{Related Work}

\textbf{Sequence-based \& Graph-based Recommendation. }
\emph{Sequence-based recommendation} mainly leverages users' sequential behavior to mine users' interests, which focuses on individual information.
Recently, various deep neural networks have been employed for sequence-based recommendation, e.g., RNN~\cite{hidasi2015session}, memory networks~\cite{chen2018sequential}, attention mechanisms~\cite{xiao2021uprec,zhou2018deep,sun2019BERT4Rec,zeng2021knowledge} and mixed models~\cite{ying2018sequential,xi2020neural}.
\emph{Graph-based recommendation} aims to use high-order interaction information contained in the graph, which is able to model the global information of user preferences.
Existing works have proved the effectiveness of GNNs in learning user and item representations \cite{wang2019neural,xie2021long}.
In this work, we exploit both individual sequence view and global graph view in MBR.

\noindent
\textbf{Multi-behavior Recommendation.}
Inspired by transfer learning~\cite{zhuang2020comprehensive,zhu2019multi,zhu2020deep}, 
multi-behavior recommendation takes advantage of other behavior of user to help the prediction of target behavior. Ajit et al.~\cite{singh2008relational} take multi-behavior into consideration via a collective matrix factorization. Recent works often model MBR via sequence or graph-based models\cite{xi2021modeling,xie2021personalized}.
MRIG \cite{wang2020beyond} builds sequential graphs on users' behavior sequences.
MBGCN~\cite{jin2020multi} learns user-item and item-item similarities on the designed user-item graph and different co-behavior graphs.
Other works combine MBR with meta-learning \cite{xia2021graph} and external knowledge \cite{xia2021knowledge}.
However, these methods do not make full use of the correlations between behaviors via CL. In this paper, we propose a universal framework that utilizes contrastive learning to model the relations of different behaviors.

\noindent
\textbf{Self-supervised Learning.}
Self-supervised learning (SSL) aims at training a network by pretext tasks, which are designed  according to the characteristics of raw data. Recently, self-supervised learning  has been shown its superior ability in CV~\cite {doersch2015unsupervised,zhang2016colorful}, NLP~\cite{devlin2018bert}, and Graph~\cite{perozzi2014deepwalk} fields. Some works also adopt self-supervised learning in recommender systems ~\cite{zhou2020s3,xie2020contrastive,wu2021self,xie2021contrastive}. 

However, most of them fall into single-behavior methods.
In this paper, we focus on modeling the commonalities and differences between multiple behaviors and views of users with CL. 

\section{Methodology}
\subsection{Preliminaries}
MMCLR aims to make full use of multi-behavior and multi-view information to learn better representations for recommendation. We first give detailed definitions of the key notions in our multi-behavior recommendation as follows:

\noindent
\textbf{Multi-behavior Modeling}.
In MBR, the most important and profitable behavior (e.g., \emph{purchase} in E-commerce) is regarded as the target behavior. While it suffers from data sparsity issues. 
Specifically, we denote the user and item as $u\in U$ and $v \in V$, where $U$ and $V$ are user set and item set. We suppose that users have $B$ types of behaviors $\{b_1, \cdots, b_{B}\}$ in a system, where $b_t$ is the target behavior.
\noindent
\textbf{Multi-view Modeling}.
Users' multiple behaviors can be modeled with different views, highlighting different aspects of user preferences. In this work, we construct two views, including the sequence vie and the graph view.
   For the \emph{\textbf{sequence view}}, 
   we represent the multi-behavior historical sequence of user $u$ as $S_{u}=\{s_{u}^{b_1}, s_{u}^{b_2}, ..., s_{u}^{b_B}\}$, where $s_{u}^b$ is the behavior sequence of user $u$ under behavior $b$. For each behavior, we have the item sequence $s^{b}_{u}=\{v_1,v_2,...,v_{|s_{u}^b|}\}$.
   For the \emph{\textbf{graph view}}, we build a global multi-relation user-item graph ${G=(\mathcal{V},\mathcal{E})}$, where $\mathcal{V}$ is the node set, and $\mathcal{E}$ is the edge set. If user $u$ and item $v$ have an interaction under a certain behavior $b$, there is a edge $e=(u,v,b) \in \mathcal{E}$ in graph $G$. We use $\bm{u}_i^0$ and $\bm{v}_j^0$ to represent the corresponding raw feature of $u_i$ and $v_j$.

\noindent
\textbf{Problem definition}. Given a user's  multi-behavior sequences $S_{u}$ and the global multi-relation user-item graph $G$, MMCLR should predict the most appropriate  item $v$ that the user will interact under the target behavior $b_t$.

\subsection{Framework of Multi-view Multi-behavior Recommendation}

\subsubsection{Overview.}

\begin{figure*}[!hbtp]
\centering
\includegraphics[width=1.0\textwidth]{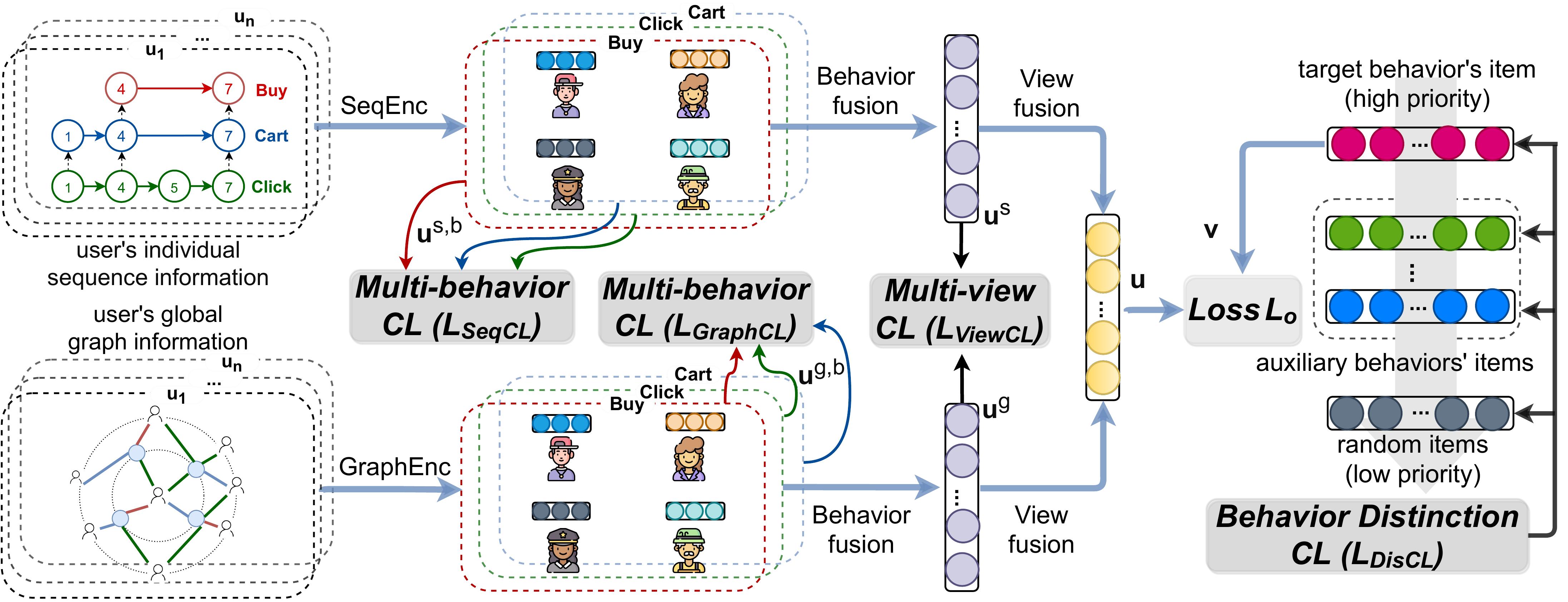}
\caption{Overall architecture of MMCLR with our proposed contrastive learning tasks.}
\label{fig:architecture}
\end{figure*}

The model structure of MMCLR is illustrated in Fig.\ref{fig:architecture}. Our model mainly has three parts: multi-view encoder, multi-behavior fusion, and multi-view fusion. Three types of contrastive learning tasks are proposed to capture the multi-behavior and multi-view feature interactions.
Specifically, for a user $u$, the global user-item graph $G$ and the user's multi-behavior sequence $S_{u}$ are first fed to the sequence-view encoder and the graph-view encoder as inputs. In both sequence and graph encoders, we build $B$ user single-behavior representations according to each behavior,  respectively.
Second, these single-behavior representations under the same view are fused by the multi-behavior fusion module, with sequence/graph-based multi-behavior CL and behavior distinction CL tasks as auxiliary losses.
Then, we combine the sequence-view and graph-view user representations by the multi-view fusion module with the multi-view CL, jointly considering  individual and global preferences.
Finally, the similarity between the fused user and item representations is viewed as the ranking score.

\subsubsection{Multi-view Encoder.}

Conventional sequence-based recommendation models \cite{zhou2018atrank,sun2019BERT4Rec} often focus on the individual historical behaviors of a user, which aims to precisely capture the local sequential information of a user. In contrast, graph-based recommendation models \cite{zheng2020price,jin2020multi} are often conducted on the whole user-item graph built by all users' behaviors, which can benefit from the global interactions. We argue that both individual sequence and global graph views are beneficial in multi-behavior recommendation.

Specifically, we implement an individual sequence-based encoder $\mathrm{SeqEnc}(\cdot)$ and a global graph-based encoder $\mathrm{GraphEnc}(\cdot)$ to learn users' and items' single-behavior representations separately. 
Formally, for the behavior $b$:
\begin{equation}
\bm{u}^{s,b} = \mathrm{SeqEnc}^b(\bm{s}^{b}_{u}), \quad
\bm{u}^{g,b} = \mathrm{GraphEnc}^b(G,u,b),
\label{eq.simple_encoder}
\end{equation}
where $\bm{s}^{b}_{u}$ is the user's historical behavior sequence of $b$, and $G$ is the global user-item graph. $\bm{u}^{s,b}$ and $\bm{u}^{g,b}$ indicate the user sequence-view and graph-view single-behavior representation of $b$. Finally, we learn $2B$ single-behavior representations in two views for the next multi-behavior and multi-view fusions. Note that we can flexibly select appropriate sequence and graph models for $\mathrm{SeqEnc}^b(\cdot)$ and $\mathrm{GraphEnc}^b(\cdot)$. Specifically,  We adopt Bert4rec and lightGCN as sequence encoder and graph encoder. For lightGCN we replace the original aggregator with meaning aggregator.

\subsubsection{Multi-behavior Fusion.}
Single-behavior representations may suffer from data sparsity issues, especially for some high-cost and low-frequent target behaviors (e.g., \emph{purchase}). In this case, other auxiliary behaviors (e.g., \emph{click}, \emph{add to cart}) could provide essential information to infer user preferences on the target behaviors. Hence, we build a multi-behavior fusion module to fuse user single-behavior representations in each view to get the integrated sequence-view representation $\bm{u}^s$ and the integrated graph-view representation $\bm{u}^g$, which is noted as:
\begin{equation}
\begin{split}
\bm{u}^s=\mathrm{MLP}^{s}(\bm{u}^{s,b_1} ||, \cdots, || \bm{u}^{s,b_{B}}), \quad
\bm{u}^g=\mathrm{MLP}^{g}(\bm{u}^0 || \bm{u}^{g,b_1} ||, \cdots, || \bm{u}^{g,b_{B}}).
\end{split}
\label{eq.user_view_representation}
\end{equation}
$\bm{u}^0$ is the raw user embedding in the graph view. $\mathrm{MLP}^s$ and $\mathrm{MLP}^g$ are two-layer MLPs with $\mathrm{ReLU}$ as activation.  We also build the graph-view item representation $\bm{v}^g$ similar to $\bm{u}^g$, where $\bm{v}^0$ is also used as the raw behavior features in Eq. (\ref{eq.simple_encoder}).

\subsubsection{Multi-view Fusion.}

To take advantage of representations in both views, we apply a multi-view fusion to learn the final user and item representations, which contain both individual and global information. We formalize the integrated user representation $\bm{u}$ and item representation $\bm{v}$ as follows:
\begin{equation}
\begin{split}
\bm{u} = \mathrm{MLP}^U (\bm{u}^s || \bm{u}^g), \quad
\bm{v} = \mathrm{MLP}^V (\bm{v}^0 || \bm{v}^g).
\end{split}
\end{equation}
Following the classical ranking model \cite{rendle2012bpr}, the inner product of $\bm{u}$ and $\bm{v}$ is used to calculate the ranking scores of user-item pairs, trained under $L_o$ as:
\begin{equation}
\begin{split}
L_o = -\sum_{(u,v_i) \in S^+} \sum_{(u,v_j) \in S^-} \log \sigma (\bm{u}^\top\bm{v}_i - \bm{u}^\top\bm{v}_j),
\end{split}
\label{eq.simple_loss}
\end{equation}
where $(u,v_i) \in S^+$ indicates the positive set of the target behavior, and $(u,v_j) \in S^-$ indicates the randomly-sampled negative set.
\subsubsection{Multi-view Multi-behavior Contrastive Learning.}

The above architecture is a straightforward combination of multi-view multi-behavior representations. To better capture the coarse-grained commonalities and fine-grained differences between different behaviors and views to learn better user representations in different views and behaviors, we design three types of CL tasks. Next we will introduce details of them.

\subsection{Multi-behavior Contrastive Learning}
\label{sec.multi-behavior_CL}

A user's single-behavior representations reflect user preferences on the corresponding behaviors, which also share certain commonalities to reflect the user itself. We build two multi-behavior CL tasks in the sequence and graph views respectively as auxiliary losses to better use multi-behavior information.

\subsubsection{Sequential Multi-behavior CL.}

We adopt a sequential multi-behavior CL, which attempts to minimize the differences between different single-behavior representations of the same user and maximize the differences between different users. In this case, we naturally regard different single-behavior representations of a user as certain kinds of (behavior-level) user augmentations.

Precisely, considering a mini-batch of $N$ users $\{u_1, \cdots, u_N\}$, we randomly select two single-behavior representations $(\bm{u}_i^{s,b_1}, \bm{u}_i^{s,b_2})$ of behavior $b_1$ and $b_2$ for each $u_i$ as the positive pair in CL. 
And we consider $(\bm{u}_i^{s,b_1}, \bm{u}_j^{s,b_2})$ as the negative pair. Following~\cite{chen2020efficient}, we also conduct a projector $\mathrm{MLP}_{p_1}(\cdot)$ to map all user single-behavior representations into the same sequential semantic space. We have:
\begin{equation}
\bm{u}_{i,p_1}^{s,b_1} = \mathrm{MLP}_{p_1} (\bm{u}_{i}^{s,b_1}), \quad
\bm{u}_{j,p_1}^{s,b_2} = \mathrm{MLP}_{p_1} (\bm{u}_{j}^{s,b_2}).
\label{eq.seq_CL_MLP}
\end{equation}
The sequential multi-behavior CL loss $L_{SeqCL}$ is defined as follows:
\begin{equation}
\begin{split}
L_{SeqCL} = - \sum_{i=1}^{N} \sum_{u_j \neq u_i} f(\bm{u}_{i,p_1}^{s,b_1},\bm{u}_{i,p_1}^{s,b_2},\bm{u}_{j,p_1}^{s,b_2}), \\
f(\bm{x},\bm{y},\bm{z}) = \log(\sigma(\bm{x}^\top\bm{y}-\bm{x}^\top\bm{z})).
\label{eq.CL_loss}
\end{split}
\end{equation}
$f(\bm{x},\bm{y},\bm{z})$ denotes our pair-wise distance function, $\sigma(\cdot)$ is the sigmoid activation.

\subsubsection{Graphic Multi-behavior CL.}

Similar with the sequential multi-behavior CL, we also build a graphic multi-behavior CL for the graph-view representations. For $\bm{u}_i^{g,b_1}$, we consider $\bm{u}_i^{g,b_2}$ as the positive sample and $\bm{u}_j^{g,b_2}$ as the negative sample in this CL. We also have $\bm{u}_{i,p_2}^{g,b_1} = \mathrm{MLP}_{p_2} (\bm{u}_{i}^{g,b_1})$ and $\bm{u}_{j,p_2}^{g,b_2} = \mathrm{MLP}_{p_2} (\bm{u}_{j}^{g,b_2})$ as Eq. (\ref{eq.seq_CL_MLP}). We define the graphic multi-behavior CL loss $L_{GraphCL}$ as follows:
\begin{equation}
\begin{split}
L_{GraphCL} = - \sum_{i=1}^{N} \sum_{u_j \neq u_i} f(\bm{u}_{i,p_2}^{g,b_1},\bm{u}_{i,p_2}^{g,b_2},\bm{u}_{j,p_2}^{g,b_2}),
\end{split}
\label{eq.L_GraphCL}
\end{equation}
in which $f(\bm{x},\bm{y},\bm{z})$ is the same as Eq. (\ref{eq.CL_loss}). Through the sequential and graphic multi-behavior CL tasks, MMCLR can learn better and more robust single-behavior representations, which is the fundamental of user diverse preferences.
It functions well, especially when the target behaviors are sparse.

\subsection{Multi-view Contrastive Learning}
\label{sec.multi-view_CL}

The multi-view CL aims to highlight the relationships between the individual sequence and global graph views. It is natural that the sequence-view and graph-view user representations of the same user should be closer than others, since they reflect the same user's preferences (though learned from different information).
Hence, we propose the multi-view CL task on the integrated sequence-view and graph-view user representations in Eq. (\ref{eq.user_view_representation}). We regard $(\bm{u}_i^{s}, \bm{u}_i^{g})$ of the same user $u_i$ as the positive pair,
considering $\bm{u}_i^{s}$ and $\bm{u}_i^{g}$ as different view-level user augmentations of $u_i$,
and regard $(\bm{u}_i^{s}, \bm{u}_j^{g})$ and $(\bm{u}_i^{g}, \bm{u}_j^{s})$ as the in-batch negative pairs of two views. After the projector, we have $\bm{u}_{i,p_3}^{s} = \mathrm{MLP}_{p_3} (\bm{u}_{i}^{s})$ and $\bm{u}_{j,p_3}^{g} = \mathrm{MLP}_{p_3} (\bm{u}_{j}^{g})$. The multi-view CL loss $L_{ViewCL}$ is noted as follows:
\begin{equation}
\begin{split}
L_{ViewCL} = - \sum_{i=1}^{N} \sum_{u_j \neq u_i} f(\bm{u}_{i,p_3}^{s},\bm{u}_{i,p_3}^{g},\bm{u}_{j,p_3}^{g}).
\end{split}\label{eq.L_ViewCL}
\end{equation}

We are the first to propose the notion of multi-view CL. Through this CL, individual sequence and global graph views can cooperate well in MBR.

\subsection{Behavior Distinction Contrastive Learning}
\label{sec.behavior_distinction_CL}


The above two CL tasks highlight the commonalities between a user's multiple behaviors and views compared to other users' representations. However, the fine-grained differences between different behaviors of a user are also essential. For example, in E-commerce, the low-frequent high-cost \emph{purchase} behaviors reflect the user's high-priority preferences, comparing with other low-cost auxiliary behaviors like \emph{click} and \emph{add to cart}. To some extent, these auxiliary behaviors (viewed as positive pair instances in multi-behavior CL) could be even regarded as certain hard negative samples of the high-cost target behaviors \cite{huang2020embedding}. Considering the fine-grained differences and behavior priorities can further improve the target behavior's (e.g., purchase) performances, especially when distinguishing ``the good but negative'' candidates (e.g., clicked but not purchased items), which are challenging interference terms in practical ranking systems. Hence, we propose a novel \textbf{behavior distinction CL} for the first time in MBR.

Specifically, we define the behavior priority in MBR as follows: \emph{items of the target behavior} $v_i$ $>$ \emph{items of auxiliary behaviors} $v_j$ $>>$ \emph{other random in-batch items} $v_k$. In the target behavior prediction task, the integrated user representation $\bm{u}$ should firstly be close to $\bm{v}_i$, and then the hard negative samples of auxiliary behaviors $\bm{v}_j$, and finally be distinct with the random negative items $\bm{v}_k$.
Similarly, we conduct a projector $\mathrm{MLP}_{p_4}$ to get $\bm{u}_{p_4}$, $\bm{v}_{i,p_4}$, $\bm{v}_{j,p_4}$, and $\bm{v}_{k,p_4}$, and then learn the item-aspect behavior distinction CL $L_{DisCL}$ as follows:

\begin{equation}
\begin{split}
L_{DisCL} = - \sum_{u} \sum_{(v_i,v_j)} \sum_{v_k} ( f(\bm{u}_{p_4},\bm{v}_{i,p_4},\bm{v}_{j,p_4}) +\beta f(\bm{u}_{p_4},\bm{v}_{j,p_4},\bm{v}_{k,p_4}) ).
\end{split}
\label{eq.L_DisCLv}
\end{equation}
$\beta$ is a loss weight, $v_i$ and $v_j$ are one of the target/auxiliary behaviors of $u$.

The multi-behavior CL (i.e., Eq. (\ref{eq.CL_loss}, \ref{eq.L_GraphCL})) aims to narrow the distances between different behaviors of a user from the global perspective, thus distinguishing them from other items. In contrast, the behavior distinction CL explores to capture the fine-grained differences between different types of behaviors of a user, achieving deeper and more precise understandings of user's target-behavior preferences.

\subsection{Optimization}
\label{sec.implementation}
\noindent
\textbf{Overall Loss.}
The overall loss $L$ is defined with hyper-parameters $\lambda$ as:
\begin{equation}
\begin{split}
L= \lambda_o L_o + \lambda_1 L_{SeqCL} + \lambda_2 L_{GraphCL} + \lambda_3 L_{ViewCL} + \lambda_4 L_{DisCL}.
\end{split}
\label{eq.L}
\end{equation}

\noindent
\textbf{Model Analysis.}

For complexity, the graph and sequential encoders can run parallel, so the encoder complexity is decided by the more complex model. Hence, MMCLR does not produce extra encoding time. For contrastive tasks, the training complexity of the MLP layer is $O(|U|d^2)$, and the complexity of CL is $O(|U|Nd)$, where $|U|$ is the number of users and $N$ is the batch size. The  complexity is equal with existing CL models \cite{zhou2020s3,wu2021self} and can be computed in parallel with fusion operations. Moreover, the CL losses are only calculated in offline, which means our model has equal online serving  complexity as others.

\section{Experiments}

In this section, we aim at answering the following research questions:
\textbf{(RQ1)} How does MMCLR perform compared with other SOTA baselines in MBR on various evaluation metrics?
\textbf{(RQ2)} What are the effects of different contrastive learning tasks in our proposed MMCLR?
\textbf{(RQ3)} How does MMCLR perform on cold-start scenarios compared to baselines and ablation versions?
\textbf{(RQ4)} How do different hyper-parameters affect the final performance?

\subsection{Datasets}

We evaluate MMCLR on two real-world MBR datasets on E-commerce, including the Tmall and CIKM2019 EComm AI dataset. \textit{Tmall\footnote{https://tianchi.aliyun.com/competition/entrance/231721/introduction}:} It is collected by Tmall, which is one of the largest E-commerce platforms in China. We process this dataset following~\cite{chen2020efficient}. After processing, our Tmall dataset contains 22,014 users and 27,155 items. We consider three behaviors (i.e., click, add-to-cart, purchase), collecting 83,778 purchase behaviors, 44,717 add-to-cart behaviors, and 485,483 click behaviors. \textit{CIKM2019 EComm AI:} It is provided by the CIKM2019 EComm AI challenge. In this dataset, each instance is made up by an item, a user and a behavior label (i.e., click, add-to-cart, purchase). 
We process this dataset following~\cite{chen2020efficient} as well. Finally, this dataset includes 23,032 users, 25,054 items, 100,529 purchase behaviors, 38,347 add-to-cart behaviors, and 276,750 click behaviors.

\subsection{Competitors}

We compare MMCLR against several state-of-the-art baselines. For baselines not designed for MBR, we adopt our MMCLR's fusion function to jointly consider multi-behavior data. All baselines exploit data of multiple behaviors.

\begin{itemize}
    \item \textbf{BERT4Rec$^{MB}$.} BERT4Rec \cite{sun2019BERT4Rec} is a self-attention-based sequential recommendation model. We conduct separate Transformer encoders on all behaviors, and fuse them via MMCLR's fusion function, denoted as BERT4Rec$^{MB}$.
    \item \textbf{LightGCN$^{MB}$.} lightGCN \cite{he2020lightgcn} is a widely-used GNN model. Similarly, we construct multiple user-item graphs for all behaviors, encode them by it.
    \item \textbf{MRIG.} MRIG \cite{wang2020beyond} is one of the SOTA sequence-based models for MBR. It adopts user's individual behavior sequence to build a sequential graph, which regards two items having an edge if they are adjacent in a sequence.
    \item \textbf{MBGCN.} MBGCN \cite{jin2020multi} is a recent graph-based MBR model. It integrates multi-behavior information by user-item and item-item propagations.
    \item \textbf{MBGMN.} MBGMN \cite{xia2021graph} is one of the SOTA graph-based models for MBR. MBGMN first models the behavior heterogeneity and interaction diversity jointly with the meta-learning paradigm.
    \item \textbf{MGNN.} MGNN \cite{zhang2020multiplex} is one of the SOTA multiplex-graph-based models for MBR. It builds users'  multi-behavior to a multiplex-graph and learns shared graph embedding and behavior-specific embedding for recommendation.
\end{itemize}
We also compare with MMCLR's ablation versions for further comparisons:
\begin{itemize}
     \item \textbf{BERT4Rec$^{CL}$.} We add the sequential multi-behavior CL $L_{SeqCL}$ to the BERT4Rec$^{MB}$, which is noted as BERT4Rec$^{CL}$.
     \item \textbf{LightGCN$^{CL}$.} Similarly, We also add the graphic multi-behavior CL to the LightGCN$^{MB}$, which is denoted as LightGCN$^{CL}$.
    \item\textbf{MMR.} MMR is an ablation version of MMCLR without all CL tasks. It can be viewed as a simple multi-view multi-behavior model, which combines BERT4Rec$^{MB}$ with LightGCN$^{MB}$ via embedding concatenation and MLP.
\end{itemize}


\subsection{Experimental Settings}

\textbf{Parameter Settings.}
The embedding sizes of users and items are $64$ and batch size is 256 for all methods. We optimize all models by Adam optimizer.
For BERT4Rec, we stack two-layer transformers and each transformer with two attention heads. The depth of our graph encoder is set to $2$. The learning rate and L2 normalization coefficient of MMCLR are set as $1e^{-3}$ and $1e^{-4}$, respectively. The weights of supervised loss $L_o$ and four CL losses (i.e., $L_{SeqCL}$, $L_{GraphCL}$, $L_{ViewCL}$, $L_{DisCL}$) are set as $1.0$, $0.2$, $0.2$, $0.2$, and $0.05$, respectively. For all baselines, We conduct a grid search for parameter selections.


\textbf{Evaluation Protocols.}
Following~\cite{xie2020contrastive,zhou2020s3}, We adopt the leave-one-out strategy to evaluate the models' performance;  We also employ the top-K hit rate (HIT), top-K Normalized Discounted Cumulative Gain (NDCG), Mean Reciprocal Rank (MRR), and AUC (Area Under the Curve). For HIT and NDCG, we report top 5 and 10; For each ground truth, we randomly sample $99$ items that user did not interact with under the target behavior as negative samples.

\subsection{Results of Multi-behavior Recommendation (RQ1)}

The main MBR results are shown in Table \ref{table_3}, from which we find that:

(1) MMCLR performs the best among all baselines and ablation versions of MMCLR on all metrics in two datasets. It achieves $4\% \sim 11.8\%$  improvements over the best baselines on most metrics, with the significance level as $p<0.05$ (paired t-test of MMCLR V.S. baselines). It indicates that MMCLR can well capture the commonalities and differences between different behaviors and views, and thus can better take advantage of all multi-view and multi-behavior information in MBR.(2) BERT4Rec$^{CL}$ and LightGCN$^{CL}$ perform much better than their original models without CL. It verifies the importance of modeling relations between different types of behaviors when jointly learning user representations. It also implies that our multi-behavior CL can help to capture the behavior-level commonalities. Nevertheless, MMCLR still performs better than single-view models, which verifies the significance of jointly modeling multi-view information.(3) We notice that MMR performs comparably with BERT4Rec$^{MB}$. It reflects that the simple fusion of individual sequence-based and global graph-based models may not make full use of the multi-view information. 
\begin{table}
\setlength{\abovecaptionskip}{0pt}
\setlength{\belowcaptionskip}{0pt}

\caption{Results on multi-behavior recommendation. * indicates significance (p$<$0.05).}
\label{table_3}
\center
\small
\begin{tabular}{l|l|cccccc}
\toprule

Database & Model &MRR\quad&AUC\quad& HIT@5\quad& NDCG@5\quad&  HIT@10\quad& NDCG@10 \\
\midrule
\multirow{11}{*}{Tmall}
~  &BERT4Rec$^{MB}$&  0.1568& 0.6671 &0.2138& 0.1448 &0.3133&0.1769 \\
~  &LightGCN$^{MB}$& 0.1449   &  0.6542& 0.1983 & 0.1318 &0.3020 &0.1651\\
~  &MRIG&            0.1545&    0.6823&0.2084 &0.1401 &0.3207&0.1762\\
~  &MBGCN & 0.1534& 0.6912 & 0.2100& 0.1396 &0.3208&0.1751 \\
~ &MBGMN &0.1673 &0.6808 &0.2273 &0.1559 &0.3308 &0.1892\\
~  &MGNN& \underline{0.1782}& \underline{0.6955} & \underline{0.2332}& \underline{0.1651} &\underline{0.3389}&\underline{0.1991} \\
\cmidrule{2-8}
~  &LightGCN$^{CL}$ & 0.1609& 0.6863 & 0.2201& {0.1483} &0.3293&0.1835 \\
~  &BERT4Rec$^{CL}$ & 0.1754& 0.6971 & 0.2385& 0.1641 &0.3467&0.1990 \\
~  &MMR& 0.1576 & 0.6606 & 0.2152  & {0.1466} &0.3108&0.1773\\
\cmidrule{2-8}
~  &MMCLR & \textbf{0.1861*} &\textbf{0.7237*} & \textbf{0.2608*}  &  \textbf{0.1770*} &\textbf{0.3751*}&\textbf{0.2138*} \\
~  &Improvement& \textbf{4.4\%} & \textbf{4.1\%} & \textbf{11.8\%}  & \textbf{7.3\%} &\textbf{10.7\%}&\textbf{7.4\%}\\
\bottomrule
\toprule
\multirow{11}{*}{CIKM}
~  &BERT4Rec$^{MB}$&  0.1792& 0.6990 &0.2451& 0.1687 &0.3552&0.2042 \\
~  &LightGCN$^{MB}$& 0.1705   &  0.6979& 0.2332 & 0.1584 &0.3466 &0.1949\\
~  &MRIG&            0.1795&    0.7026&0.2489 &0.1696 &0.3649&0.2068\\
~  &MBGCN & 0.1850& 0.6897 &0.2479& 0.1751 &0.3492&0.2077 \\
~ &MBGMN &0.1887 &0.7035 &0.2575 &0.1795 &0.3648 &0.2140\\
~  &MGNN& \underline{0.1973}& \underline{0.7116} & \underline{0.2616}& \underline{0.1866} & \underline{0.3718}& \underline{0.2222}\\
\cmidrule{2-8}
~  &LightGCN$^{CL}$ & 0.1746& 0.7031 & 0.2398& {0.1633} &0.3530&0.1998 \\
~  &BERT4Rec$^{CL}$ & 0.1984& 0.7282 & 0.2728& 0.1912 &0.3929&0.2281 \\
~  &MMR& 0.1788& 0.6941 & 0.2506  & {0.1700} &0.3627&0.2061\\
\cmidrule{2-8}
~  &MMCLR & \textbf{0.2046*} &\textbf{0.7313*} & \textbf{0.2878*}  &  \textbf{0.1981*} &\textbf{0.4049*}&\textbf{0.2358*} \\
~  &Improvement& \textbf{3.7\%} &\textbf{2.9\%} & \textbf{10.0\%}  & \textbf{6.2\%} &\textbf{8.9\%}&\textbf{6.1\%} \\
\bottomrule
\end{tabular}
\end{table}

\subsection{Ablation Study (RQ2)}
\label{sec.ablation}
In this section, we aim to prove that MMCLR can solve the three challenges mentioned in the introduction section via three CL tasks. We build seven ablation versions of MMCLR, which are different combinations of CL tasks and the multi-view fusion, to show the effectiveness of different components.
Specifically, we regard the basic sequence-based model of MMCLR with multi-behavior information as \emph{seq} (i.e., BERT4Rec$^{MB}$), and the basic graph-based model of enhanced LightGCN with multi-behavior information as \emph{graph} (i.e., LightGCN$^{MB}$). We set \emph{seq+graph} as the simple multi-view fusion version (i.e., MMR). Moreover, we represent the multi-behavior CL, multi-view CL, and behavior distinction CL as BCL, VCL, and DCL, respectively. The final MMCLR is noted as seq+graph
\noindent
+BCL+VCL+DCL. From Table \ref{tab.ablation}, we can observe that:

(1) Comparing ablation versions with and without BCL, we find that both sequential and graphic multi-behavior CL tasks are beneficial. BCL tasks even function well on the seq+graph model. The improvements of BCL are impressive, which have over $2\%$ improvements in most metrics. 
It is because that multiple behaviors produced by the same user should reflect related preferences of the user. Modeling the coarse-grained commonalities of different behaviors helps to learn better representations to fight against the data sparsity issues. Moreover, through BCL, we can learn better user representations that are more precise and distinguishable from other users'. It reconfirms the effectiveness of the multi-behavior CL in modeling such coarse-grained commonality.(2) Comparing models with and without VCL, we know that the multi-view CL is also essential in multi-view fusion (getting nearly $1\%$ improvements on most metrics). We also implement a simple fusion model with seq and graph models, whose improvements over single-view models are marginal.
The multi-view CL smartly aligns sequence-view and graph-view representations via the CL-based learning, which well captures useful information from both individual and global aspects. These improvements verify the significance of multi-view CL.(3) Comparing with the last two versions, we can observe that the behavior distinction CL further improves the performances on all metrics. The $0.6-1.4\%$ improvements are significant. It verifies that jointly considering both coarse-grained commonalities and fine-grained differences are essential in MMCLR.

\begin{table}
\setlength{\abovecaptionskip}{0pt}
\setlength{\belowcaptionskip}{0pt}

\caption{Ablation tests on CL tasks and multi-view fusion in MMCLR.}
\label{tab.ablation}
\centering
\small
\begin{tabular}{l|c|c|c|c}
\toprule

Ablation & HIT@5 & NDCG@5 & HIT@10 & NDCG@10  \\
\midrule
seq&0.2138 &0.1448&0.3133 &0.1769    \\
graph &0.2108&0.1442& 0.3136&0.1773\\
seq+graph &0.2152&0.1466&0.3108&0.1773 \\
seq+BCL&0.2385 &0.1641&0.3467 &0.1990    \\
graph+BCL &0.2380&0.1620&0.3456&0.1966 \\
seq+graph+BCL &0.2418 &0.1632&0.3527 &0.1988   \\
seq+graph+BCL+VCL &0.2521 &0.1722&0.3614 &0.2074   \\
\midrule
MMCLR (final) & \textbf{0.2608*} & \textbf{0.1770*} & \textbf{0.3751*} & \textbf{0.2138*}   \\
\bottomrule
\end{tabular}
\end{table}
\subsection{Results on Cold-start Scenarios (RQ3)}

Real-world multi-behavior recommendation systems usually suffer from cold-start issues (e.g., cold-start users that have few historical behaviors), especially for the high-cost purchase behaviors in MBR of E-commerce. Hence, we further conduct an evaluation on the cold-start (user) scenario to verify the effectiveness of MMCLR on more challenging tasks.
Without loss of generality, we regard all users that have less than $3$ target behaviors in the train set as our cold-start users and select these cold-start users' test instances in the overall Tmall dataset as the test set of the cold-start scenario.
To comprehensively display the effectiveness of MMCLR and its multiple CL tasks on the cold-start scenario, we draw three figures in Fig. \ref{fig. cold_start} from different aspects. Precisely, we can observe that:
\begin{figure}
\setlength{\abovecaptionskip}{0pt}
\setlength{\belowcaptionskip}{0pt}
\centering
\includegraphics[width=1.0\textwidth]{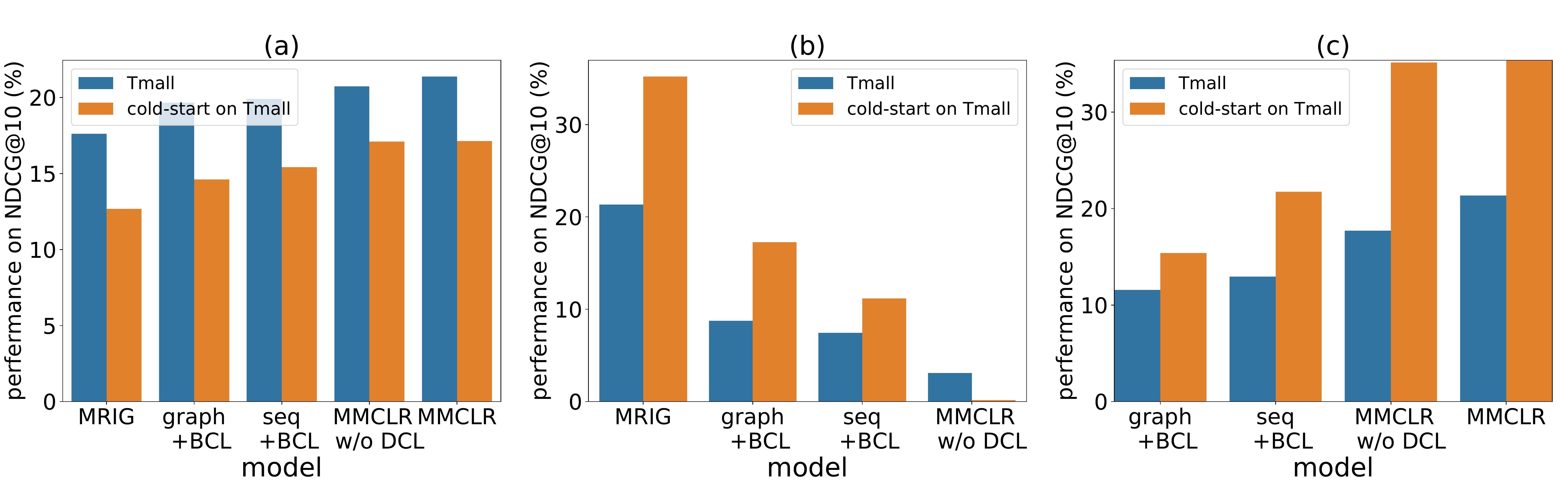}
\caption{Results of different models and ablation versions on the overall and cold-start scenarios. (a) NDCG@10 on the overall and cold-start datasets. (b) MMCLR's relative improvements of NDCG@10 on different baselines. (c) Different MMCLR's ablation versions' relative improvements of NDCG@10 on the baseline MRIG.}
\label{fig. cold_start}
\end{figure}
(1) Fig. \ref{fig. cold_start}(a) shows different models' NDCG performances in both overall and cold-start users. We can know that: (a) All models perform better on the overall users than the cold-start users. (b) Results on both overall and cold-start users have consistent improvements from graph+BCL to MMCLR.(2) Fig. \ref{fig. cold_start}(b) shows MMCLR's relative improvements on other models. We find that: (a) Comparing with different models and ablation versions (except MMCLR w/o DCL), MMCLR has higher improvements on cold-start scenarios (e.g., nearly $35\%$ astonishing improvements on MRIG). It is because that MMCLR can make full use of the multi-behavior and multi-view information via CL tasks, which can alleviate the data sparsity in cold-start users. (b) We notice that DCL brings in a slight improvement on cold-start users. It is natural since cold-start users usually have very few target behaviors, and rely more on auxiliary behaviors via the commonality-led CL tasks as supplements.(3) Fig. \ref{fig. cold_start}(c) gives the relative improvements of different MMCLR's ablation versions on MRIG. We observe that: (a) Both sequential and graphic multi-behavior CL, multi-view CL, and behavior distinction CL has improvements on cold-start scenarios. (b) Relatively, the multi-behavior CL contributes more on the overall dataset, while the multi-view CL focuses more on the cold-start users. It may be because that a different view can bring in more information for cold-start users thanks to the global graph view and its multi-view CL task.

\subsection{Parameter Analyses (RQ4)}

\begin{figure}
\setlength{\abovecaptionskip}{0pt}
\setlength{\belowcaptionskip}{0pt}
\centering
\includegraphics[width=0.9\textwidth]{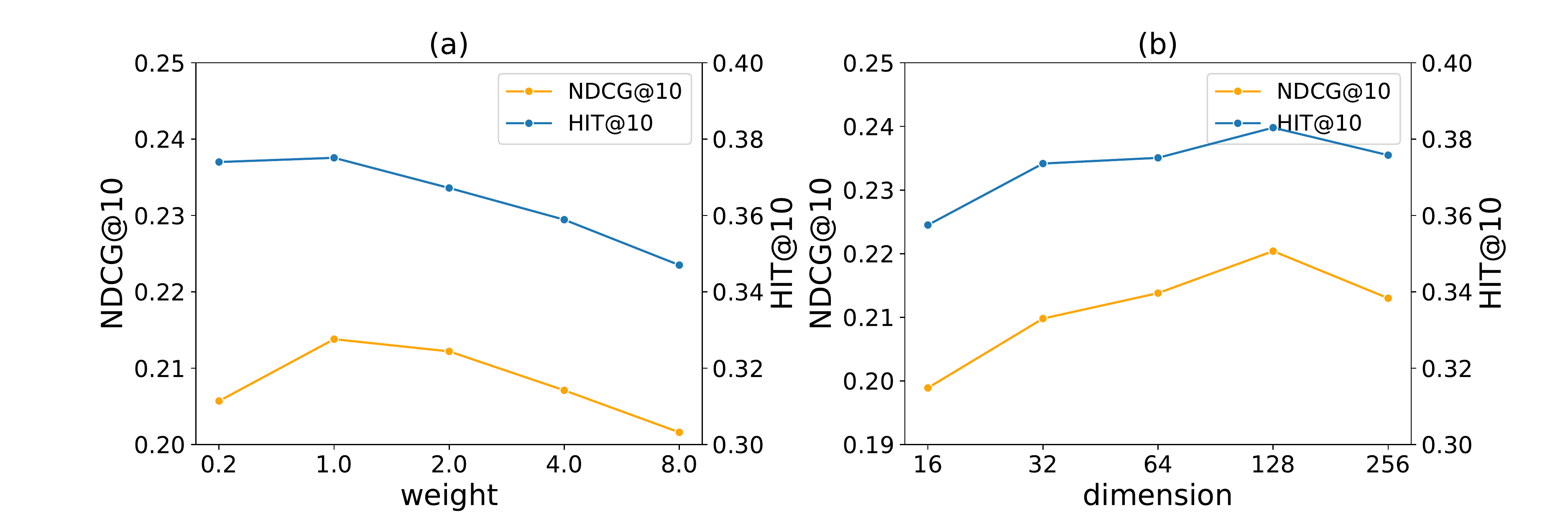}
\caption{Parameter analyses on (a) loss weights, and (b) embedding dimensions.}
\label{fig. parameter_anslyses}
\end{figure}

\noindent
\textbf{Loss Weight.}
We start the experiment with different main-task loss weights on the Tmall dataset to explore its influence. We change the weight of supervised $L_o$ among $\{0.2,1,2,4,8\}$. From Fig. \ref{fig. parameter_anslyses}(a) we can find that:
(1) Both HIT@10 and NDCG@10 first increase and then decrease from $0.2$ to $8$, and MMCLR achieves the best results when $\lambda_o=1.0$ (here CL loss weights are $0.2$, $0.2$, $0.2$, and $0.05$). It indicates that the supervised loss is the fundamental of model training, and a proper loss weight helps to balance the supervised and self-supervised learning.
(2) MMCLR consistently outperforms baselines with different weights. It shows the effectiveness and robustness of our model with different loss weights.
\noindent
\textbf{Embedding Dimension.}
We also test different input embedding dimensions on the Tmall dataset. We vary the embedding dimensions in $\{16,32,64,128,256\}$, and keep other optimal hyper-parameters unchanged. The results of different dimensions are shown in Fig. \ref{fig. parameter_anslyses}. We observe that the model achieves better performance with bigger dimension, while dimension from $16$ to $128$. It shows that enough embedding dimension helps to increase model capacity. In contrast, the model with $256$ dimensions has worse performance than $128$ dimensions. The performance may be suffered from overfitting. It also suggests that too large an embedding dimension is not necessary.

\section{Conclusion}
In this work,
We study the multi-behavior recommendation problem. Specifically, to alleviate the sparsity problem of target behaviors existing in recommender systems,
we propose a novel MMCLR framework to jointly consider the commonalities and differences between different behaviors and views in MBR via three CL tasks. 
Extensive experimental results verify the effectiveness of our MMCLR and its CL tasks.  The performance of MMCLR on cold-start users further demonstrates the superiority of MMCLR on the cold-start problem.

\section{Acknowledgments}
The research work supported by the National Natural Science Foundation of China under Grant No.61976204, U1811461, U1836206. Xiang Ao is also supported by the Project of Youth Innovation Promotion Association CAS, Beijing Nova Program Z201100006820062.

\bibliographystyle{splncs03}
\bibliography{samplepaper}

\begin{thebibliography}{10}
\providecommand{\url}[1]{\texttt{#1}}
\providecommand{\urlprefix}{URL }
\providecommand{\doi}[1]{https://doi.org/#1}

\bibitem{chen2021graph}
Chen, C., Ma, W., Zhang, M., Wang, Z., He, X., Wang, C., Liu, Y., Ma, S.: Graph
  heterogeneous multi-relational recommendation. In: Proceedings of AAAI (2021)

\bibitem{chen2020efficient}
Chen, C., Zhang, M., Zhang, Y., Ma, W., Liu, Y., Ma, S.: Efficient
  heterogeneous collaborative filtering without negative sampling for
  recommendation. In: Proceedings of AAAI (2020)

\bibitem{chen2018sequential}
Chen, X., Xu, H., Zhang, Y., Tang, J., Cao, Y., Qin, Z., Zha, H.: Sequential
  recommendation with user memory networks. In: Proceedings of WSDM (2018)

\bibitem{devlin2018bert}
Devlin, J., Chang, M.W., Lee, K., Toutanova, K.: Bert: Pre-training of deep
  bidirectional transformers for language understanding. arXiv preprint  (2018)

\bibitem{doersch2015unsupervised}
Doersch, C., Gupta, A., Efros, A.A.: Unsupervised visual representation
  learning by context prediction. In: Proceedings of ICCV (2015)

\bibitem{he2020lightgcn}
He, X., Deng, K., Wang, X., Li, Y., Zhang, Y., Wang, M.: Lightgcn: Simplifying
  and powering graph convolution network for recommendation. In: Proceedings of
  SIGIR (2020)

\bibitem{hidasi2015session}
Hidasi, B., Karatzoglou, A., Baltrunas, L., Tikk, D.: Session-based
  recommendations with recurrent neural networks. In: ICLR (2016)

\bibitem{huang2020embedding}
Huang, J.T., Sharma, A., Sun, S., Xia, L., Zhang, D., Pronin, P., Padmanabhan,
  J., Ottaviano, G., Yang, L.: Embedding-based retrieval in facebook search.
  In: Proceedings of KDD (2020)

\bibitem{jin2020multi}
Jin, B., Gao, C., He, X., Jin, D., Li, Y.: Multi-behavior recommendation with
  graph convolutional networks. In: Proceedings of SIGIR (2020)

\bibitem{pan2019warm}
Pan, F., Li, S., Ao, X., Tang, P., He, Q.: Warm up cold-start advertisements:
  Improving ctr predictions via learning to learn id embeddings. In:
  Proceedings of SIGIR. pp. 695--704 (2019)

\bibitem{pan2010transfer}
Pan, W., Xiang, E., Liu, N., Yang, Q.: Transfer learning in collaborative
  filtering for sparsity reduction. In: Proceedings of AAAI. vol.~24 (2010)

\bibitem{perozzi2014deepwalk}
Perozzi, B., Al-Rfou, R., Skiena, S.: Deepwalk: Online learning of social
  representations. In: Proceedings of KDD (2014)

\bibitem{rendle2012bpr}
Rendle, S., Freudenthaler, C., Gantner, Z., Schmidt-Thieme, L.: Bpr: Bayesian
  personalized ranking from implicit feedback. arXiv preprint  (2012)

\bibitem{singh2008relational}
Singh, A.P., Gordon, G.J.: Relational learning via collective matrix
  factorization. In: Proceedings of KDD. pp. 650--658 (2008)

\bibitem{sun2019BERT4Rec}
Sun, F., Liu, J., Wu, J., Pei, C., Lin, X., Ou, W., Jiang, P.: Bert4rec:
  Sequential recommendation with bidirectional encoder representations from
  transformer. In: Proceedings of CIKM (2019)

\bibitem{wang2020beyond}
Wang, W., Zhang, W., Liu, S., Liu, Q., Zhang, B., Lin, L., Zha, H.: Beyond
  clicks: Modeling multi-relational item graph for session-based target
  behavior prediction. In: Proceedings of WWW (2020)

\bibitem{wang2019neural}
Wang, X., He, X., Wang, M., Feng, F., Chua, T.S.: Neural graph collaborative
  filtering. In: Proceedings of SIGIR (2019)

\bibitem{wu2021self}
Wu, J., Wang, X., Feng, F., He, X., Chen, L., Lian, J., Xie, X.:
  Self-supervised graph learning for recommendation. In: Proceedings of SIGIR
  (2021)

\bibitem{xi2021modeling}
Xi, D., Chen, Z., Yan, P., Zhang, Y., Zhu, Y., Zhuang, F., Chen, Y.: Modeling
  the sequential dependence among audience multi-step conversions with
  multi-task learning in targeted display advertising. In: Proceedings of KDD
  (2021)

\bibitem{xi2020neural}
Xi, D., Zhuang, F., Song, B., Zhu, Y., Chen, S., Hong, D., Chen, T., Gu, X.,
  He, Q.: Neural hierarchical factorization machines for user's event sequence
  analysis. In: Proceedings of SIGIR. pp. 1893--1896 (2020)

\bibitem{xia2021knowledge}
Xia, L., Huang, C., Xu, Y., Dai, P., Zhang, X., Yang, H., Pei, J., Bo, L.:
  Knowledge-enhanced hierarchical graph transformer network for multi-behavior
  recommendation. In: Proceedings of AAAI (2021)

\bibitem{xia2021graph}
Xia, L., Xu, Y., Huang, C., Dai, P., Bo, L.: Graph meta network for
  multi-behavior recommendation. In: Proceedings of SIGIR (2021)

\bibitem{xiao2021uprec}
Xiao, C., Xie, R., Yao, Y., Liu, Z., Sun, M., Zhang, X., Lin, L.: Uprec:
  User-aware pre-training for recommender systems. arXiv preprint  (2021)

\bibitem{xie2021contrastive}
Xie, R., Liu, Q., Wang, L., Liu, S., Zhang, B., Lin, L.: Contrastive
  cross-domain recommendation in matching (2021)

\bibitem{xie2021personalized}
Xie, R., Liu, Y., Zhang, S., Wang, R., Xia, F., Lin, L.: Personalized
  approximate pareto-efficient recommendation. In: Proceedings of the Web
  Conference 2021. pp. 3839--3849 (2021)

\bibitem{xie2020internal}
Xie, R., Qiu, Z., Rao, J., Liu, Y., Zhang, B., Lin, L.: Internal and contextual
  attention network for cold-start multi-channel matching in recommendation.
  In: Proceedings of IJCAI. pp. 2732--2738 (2020)

\bibitem{xie2021long}
Xie, R., Wang, Y., Wang, R., Lu, Y., Zou, Y., Xia, F., Lin, L.: Long short-term
  temporal meta-learning in online recommendation. In: Proceedings of WSDM
  (2022)

\bibitem{xie2020contrastive}
Xie, X., Sun, F., Liu, Z., Wu, S., Gao, J., Ding, B., Cui, B.: Contrastive
  learning for sequential recommendation. arXiv preprint  (2020)

\bibitem{ying2018sequential}
Ying, H., Zhuang, F., Zhang, F., Liu, Y., Xu, G., Xie, X., Xiong, H., Wu, J.:
  Sequential recommender system based on hierarchical attention network. In:
  Proceedings of IJCAI (2018)

\bibitem{zeng2021knowledge}
Zeng, Z., Xiao, C., Yao, Y., Xie, R., Liu, Z., Lin, F., Lin, L., Sun, M.:
  Knowledge transfer via pre-training for recommendation: A review and
  prospect. Frontiers in big Data  (2021)

\bibitem{zhang2016colorful}
Zhang, R., Isola, P., Efros, A.A.: Colorful image colorization. In: Proceedings
  of ECCV (2016)

\bibitem{zhang2020multiplex}
Zhang, W., Mao, J., Cao, Y., Xu, C.: Multiplex graph neural networks for
  multi-behavior recommendation. In: {Proceedings of CIKM} (2020)

\bibitem{zheng2020price}
Zheng, Y., Gao, C., He, X., Li, Y., Jin, D.: Price-aware recommendation with
  graph convolutional networks. In: Proceedings of ICDE (2020)

\bibitem{zhou2018atrank}
Zhou, C., Bai, J., Song, J., Liu, X., Zhao, Z., Chen, X., Gao, J.: Atrank: An
  attention-based user behavior modeling framework for recommendation. In:
  Proceedings of AAAI (2018)

\bibitem{zhou2018deep}
Zhou, G., Zhu, X., Song, C., Fan, Y., Zhu, H., Ma, X., Yan, Y., Jin, J., Li,
  H., Gai, K.: Deep interest network for click-through rate prediction. In:
  Proceedings of KDD (2018)

\bibitem{zhou2020s3}
Zhou, K., Wang, H., Zhao, W.X., Zhu, Y., Wang, S., Zhang, F., Wang, Z., Wen,
  J.R.: S3-rec: Self-supervised learning for sequential recommendation with
  mutual information maximization. In: Proceedings of CIKM (2020)

\bibitem{zhu2021transfer}
Zhu, Y., Ge, K., Zhuang, F., Xie, R., Xi, D., Zhang, X., Lin, L., He, Q.:
  Transfer-meta framework for cross-domain recommendation to cold-start users.
  In: Proceedings of SIGIR (2021)

\bibitem{zhu2021personalized}
Zhu, Y., Tang, Z., Liu, Y., Zhuang, F., Xie, R., Zhang, X., Lin, L., He, Q.:
  Personalized transfer of user preferences for cross-domain recommendation.
  In: Proceedings of WSDM (2021)

\bibitem{zhu2021learning}
Zhu, Y., Xie, R., Zhuang, F., Ge, K., Sun, Y., Zhang, X., Lin, L., Cao, J.:
  Learning to warm up cold item embeddings for cold-start recommendation with
  meta scaling and shifting networks. In: Proceedings of SIGIR (2021)

\bibitem{zhu2019multi}
Zhu, Y., Zhuang, F., Wang, J., Chen, J., Shi, Z., Wu, W., He, Q.:
  Multi-representation adaptation network for cross-domain image
  classification. Neural Networks  \textbf{119},  214--221 (2019)

\bibitem{zhu2020deep}
Zhu, Y., Zhuang, F., Wang, J., Ke, G., Chen, J., Bian, J., Xiong, H., He, Q.:
  Deep subdomain adaptation network for image classification. TNNLS  (2020)

\bibitem{zhuang2020comprehensive}
Zhuang, F., Qi, Z., Duan, K., Xi, D., Zhu, Y., Zhu, H., Xiong, H., He, Q.: A
  comprehensive survey on transfer learning. Proceedings of the IEEE  (2020)

\end{thebibliography}

\end{document}